# Modeling Waveform Shapes with Random Effects Segmental Hidden Markov Models

**Seyoung Kim, Padhraic Smyth**
Department of Computer Science
University of California, Irvine
CA 92697-3425
{sykim,smyth}@ics.uci.edu

**Stefan Luther**
Department of Applied Physics
University of Twente
The Netherlands
s.luther@tn.utwente.nl

## Abstract

In this paper we describe a general probabilistic framework for modeling waveforms such as heartbeats from ECG data. The model is based on segmental hidden Markov models (as used in speech recognition) with the addition of random effects to the generative model. The random effects component of the model handles shape variability across different waveforms within a general class of waveforms of similar shape. We show that this probabilistic model provides a unified framework for learning these models from sets of waveform data as well as parsing, classification, and prediction of new waveforms. We derive a computationally efficient EM algorithm to fit the model on multiple waveforms, and introduce a scoring method that evaluates a test waveform based on its shape. Results on two real-world data sets demonstrate that the random effects methodology leads to improved accuracy (compared to alternative approaches) on classification and segmentation of real-world waveforms.

## 1 Introduction

Automatically parsing and recognizing waveforms based on their shape is a classic problem in pattern recognition (Fu, 1982). Applications include automated classification of heartbeat waveforms in ECG data analysis (Koski, 1986), interpretation of waveforms from turbulent flow experiments (Bruun, 1995), and discrimination of nuclear events and earthquakes in seismograph data (Bennett & Murphy, 1986). Typically in these applications it is impractical for a human to continuously monitor the time-series data in real-time (or to scan large archives of such data) and there is a need for accurate and automated real-time waveform detection. Other applications of waveform modeling occur in database systems and information retrieval, for systems that can take a waveform as an "input query" and search a large database to find similar waveforms that match this query (e.g., Yi & Faloutsos, 2000). While the human visual system can easily recognize the characteristic signature of a particular waveform shape (a heartbeat waveform for example) the problem can be quite difficult for automated methods. For example, as Figure 1(a) shows, there can be significant variability in shape among waveforms belonging to the same general class.

A generally useful approach to these problems is to construct a generative model for the waveform and then use this model to detect and parse new waveforms. For example, syntactic grammars decompose the waveform into a set of component parts generated by a set of grammatical rules. To model shape variability these grammars require the addition of a stochastic component, and to learn such models from data requires a likelihood function expressing the probability of an observed set of waveforms given a model and its parameters. In this general context relatively simple statistical grammars such as hidden Markov models (HMMs) have been pursued (e.g., Koski, 1996; Hughes et al., 2003), given that stochastic grammars with richer representations are generally much more difficult to learn. The parameters of these models can be learned from a set of examplar waveforms—new waveforms can then be parsed and classified based on the likelihood of the new waveform given the trained model.

A potentially useful extension of standard HMMs for shape modeling is the so-called segmental hidden Markov model, originally introduced in the speech recognition community (Levinson, 1986; Ostendorf et al., 1996) and proposed for more general waveform modeling in Ge & Smyth (2000). The segmental model allows for the observed data within each segment (a sequence of states with the same value) to follow a general parametric regression form, such as a linear function of time with additive noise. This allows us to model the shape of the waveform directly, in this case as a sequence of piecewise linear components, as shown in Figure 1(c).

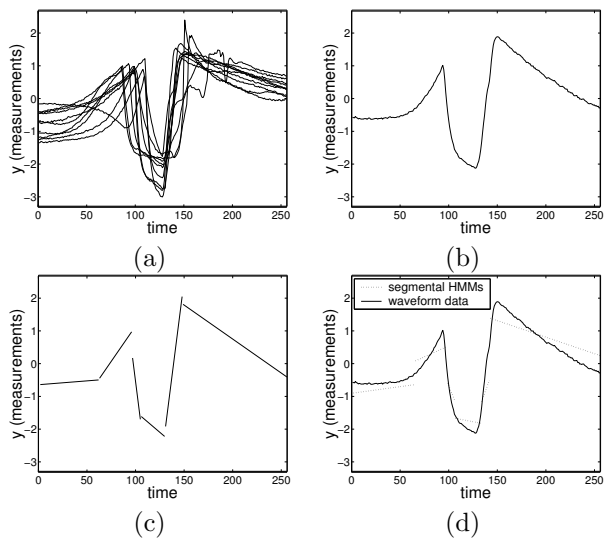

Figure 1: Bubble-probe interaction data: (a) a set of waveforms obtained from bubbles that are split by a probe during interaction, (b) an example of single waveform, (c) piecewise linear approximation of (b), and (d) segmental HMMs fit to a test waveform.

A limitation of the standard segmental model is that it assumes that the parameters of the model are fixed. Thus, the only source of variability in an observed waveform arises from variation in the lengths of the segments and observation noise added to the functional form in each segment. The limitation of this can clearly be seen in Figure 1(d), where a segmental model has been trained on the data in Figure 1(a) and then used to parse the specific waveform in Figure 1(b) ("parsing" means inferring the most likely state sequence given the model). We can see that the slopes and intercepts provided by the model do not match the observed data particularly well in each segment, e.g., in the first segment the intercept is clearly too low on the $y$-axis, in the second segment the slope is too small, and so forth. By using the same fixed parameters for all waveforms, the model cannot fully account for variability in waveform shapes.

To overcome this limitation, in this paper we combine segmental HMMs with random effects models (Laird & Ware, 1982). The general idea of modeling with random effects is to allow parameters to have individual-level (or waveform-level) random variation, while still being coupled together by an overall "population prior." By extending the segmental HMM to include random effects, we can allow the slopes and intercepts of each waveform to vary according to a prior distribution, within each segment. The parameters of this prior can be learned from data in the form of sets of waveforms in an unsupervised manner. In fact the resulting model can be viewed as a directed graphical model, allowing for application of standard methods for inference and learning. For example, we can in principle learn that the slopes across multiple waveforms for the first segment in Figure 1(c) tend to have a characteristic mean slope and standard deviation. The random effects approach provides a systematic mechanism for allowing variation in "shape space" in a manner that can be parametrized.

The primary contributions of this paper are to (a) propose the use of random effects segmental HMMs for general waveform modeling applications, (b) derive a computationally efficient EM procedure for learning such models (reducing complexity by a factor of $T^2$ where $T$ is the length of a waveform), (c) propose two separate likelihood-based scores for shape and for noise (which are then shown to improve recognition accuracy over using just likelihood alone), and finally (d) illustrate on two real waveform data sets how these models can be used for waveform parsing, classification, and prediction. The closest related work is Holmes & Russell (1999) who explored a similar idea for using a distribution over parameters in segmental HMMs, in the context of speech recognition. Our work extends these ideas by deriving a provably correct EM algorithm, showing how the computational complexity of this EM algorithm can be significantly reduced, and generalizing the applicability of the method.

We begin our discussion by introducing segmental HMMs in Section 2. In Section 3, we extend this model to incorporate random effects models, and describe the EM algorithm for parameter estimation as well as the inference algorithms and the scoring methods for test waveforms. In Section 4, we evaluate our model on two applications involving bubble-probe interaction data and ECG data, with conclusions in Section 5.

## 2 Segmental HMMs

Standard discrete-time finite-state HMMs impose a geometric distribution on run lengths (or segment lengths) for each state value and assume that observations are conditionally independent with a constant mean within such segments. Segmental HMMs relax these modeling constraints by allowing (a) arbitrary distributions on run lengths and (b) "segment models" (regression models) that allow the mean to be a function of time within each segment.

A segmental HMM with $M$ states is described by an

UAI 2004　　　　　　　　　　　　　　　　　　　　KIM ET AL.　　　　　　　　　　　　　　　　　　　　311$M \times M$ transition matrix, plus a duration distribution and segment distribution for each state $k$, where $k = 1, \ldots, M$. The transition matrix $\mathbf{A}$ (which is stationary in time) has entries $a_{kl}$, namely, the probability of being in state $k$ at time $t+1$ given state $l$ at time $t$. The initial state distribution can be included in $\mathbf{A}$ as transitions from state 0 to each state $k$. In waveform modeling, we typically constrain the transition matrix to allow only left-to-right transitions and no self-transitions. Thus, there is an ordering on states, each state can be visited at most once, and states can be skipped.

In this paper, we model the duration distribution of state $k$, using a Poisson distribution,

$$P(d|\boldsymbol{\theta}_{d_k}) = \frac{e^{-\lambda_k}\lambda_k^{d-1}}{(d-1)!} \qquad d = 1, 2, \ldots$$

(shifted to start at $d = 1$ to prevent a silent state). Other choices for the duration distribution could also be used. Once the process enters state $k$, a duration $d$ is drawn, and state $k$ produces a segment of observations of length $d$ from the segment distribution. In what follows we assume that the shape of waveforms can be approximated as a sequence of linear segments, and model the $r$th segment of observations of length $d$, $\mathbf{y}_r$, generated by state $k$, as a linear regression function in time,

$$\mathbf{y}_r = \mathbf{X}_r\boldsymbol{\beta}_k + \mathbf{e}_r \qquad \mathbf{e}_r \sim N_d(\mathbf{0}, \sigma^2\mathbf{I}_d), \qquad (1)$$

where $\boldsymbol{\beta}_k$ is a $2 \times 1$ vector of regression coefficients, $\mathbf{e}_r$ is a $d \times 1$ vector of Gaussian noise with variance $\sigma^2$ in each component, and $\mathbf{X}_r$ is a $d \times 2$ design matrix consisting of a column of 1's (for the intercept term) and a column of $x$ values representing the time values. Note that this model can easily be generalized to allow nonlinear polynomial functions of $x$ that are still linear in the parameters $\boldsymbol{\beta}_k$. For simplicity, $\sigma^2$ is assumed to be common across all states; again this can be relaxed. One could enforce continuity of the mean functions across segments in the probabilistic model, but this is not discussed in the present paper.

Treating the unobserved state sequences as missing, we can estimate the parameters, $\boldsymbol{\theta} = \{\mathbf{A}, \boldsymbol{\theta}_d = \{\lambda_k | k = 1, \ldots, M\}, \boldsymbol{\theta}_f = \{\boldsymbol{\beta}_k, (\sigma^2)|k = 1, \ldots, M\}\}$, using the EM algorithm with the forward-backward (F-B) algorithm as a subroutine for inference in the E step (Deng et al., 1994). The F-B algorithm for segmental HMMs, modified from that of standard HMMs to take into account the duration distribution, recursively computes

$$\alpha_t(k) = P(y_{1:t}, \text{stay in state } k \text{ ends at } t|\boldsymbol{\theta})$$
$$\alpha_t^*(k) = P(y_{1:t}, \text{stay in state } k \text{ starts at } t+1|\boldsymbol{\theta}) \quad (2)$$

in the forward pass, and

$$\beta_t(k) = P(y_{t+1:T}|\text{stay in state } k \text{ ends at } t, \boldsymbol{\theta})$$
$$\beta_t^*(k) = P(y_{t+1:T}|\text{stay in state } k \text{ starts at } t+1, \boldsymbol{\theta}) \quad (3)$$

in the backward pass, and returns the results to the M step as sufficient statistics (Rabiner & Juang, 1993).

Inference algorithms for segmental HMMs provide a natural way to evaluate the performance of the model on test data. The F-B algorithm scores a previously unseen waveform $\mathbf{y}$ by calculating the likelihood

$$p(\mathbf{y}|\boldsymbol{\theta}) = \sum_{\mathbf{s}} p(\mathbf{y}, \mathbf{s}|\boldsymbol{\theta}) = \sum_k \alpha_T(k). \qquad (4)$$

In addition, the Viterbi algorithm can be used to provide a segmentation of a waveform by computing the most likely state sequence. The addition of duration distributions in segmental HMMs increases the time complexity of both the F-B and Viterbi algorithms from $O(M^2T)$ for standard HMMs to $O(M^2T^2)$, where $T$ is the length of the waveform (i.e. the number of observations).

## 3 Segmental HMMs with Random Effects

A random effects model is a general statistical framework when the data generation process can be seen as having hierarchical structure. At each level of the generative process, the model defines a prior distribution over the individual group parameters, called random effects, of one level below. Typically, the random effects are not observable, so the EM algorithm is a popular approach to learning model parameters from the observed data (Dempster et al., 1981; Laird & Ware, 1982). By combining segmental HMMs and random effects models we can take advantage of the strength of each in waveform modeling.

### 3.1 The Model

Beginning with the segmental HMMs described in Section 2, we can extend the segment distributions of the model as follows. Consider the $r$th segment $\mathbf{y}_r^i$ of length $d$ from the $i$th individual waveform generated by state $k$. Following the discussion in Laird & Ware (1982), we describe the generative model as a two-stage process. At stage one, we model the observed data $\mathbf{y}_r^i$ as

$$\mathbf{y}_r^i = \mathbf{X}_r^i\boldsymbol{\beta}_k + \mathbf{X}_r^i\mathbf{u}_k^i + \mathbf{e}_r^i \qquad \mathbf{e}_r^i \sim N_d(\mathbf{0}, \sigma^2\mathbf{I}_d), \quad (5)$$

where $\mathbf{e}_r^i$ is the measurement noise, $\mathbf{X}_r^i$ is a $d \times 2$ design matrix for the time measurements corresponding to $\mathbf{y}_r^i$, $(\boldsymbol{\beta}_k + \mathbf{u}_k^i)$ are the regression coefficients, and $1 \leq i \leq N$ (for $N$ waveforms). $\boldsymbol{\beta}_k$ represents the mean regression parameters for segment $k$, and $\mathbf{u}_k^i$ represents the variation in regression (or shape) parameters for the $i$th individual waveform. At this stage, the individual random effects $\mathbf{u}_k^i$ as well as $\boldsymbol{\beta}_k$ and $\sigma^2$ are viewed as parameters. At the second stage, $\mathbf{u}_k^i$ is viewed as a random variable with distribution

$$\mathbf{u}_k^i \sim N_2(\mathbf{0}, \boldsymbol{\Psi}_k), \qquad (6)$$

where $\boldsymbol{\Psi}_k$ is a $2 \times 2$ covariance matrix, and $\mathbf{u}_k^i$ is independent of $\mathbf{e}_r^i$. In this setup, it can be shown that $\mathbf{y}_r^i$ and $\mathbf{u}_k^i$ have the following joint distribution:

$$\begin{pmatrix} \mathbf{y}_r^i \\ \mathbf{u}_k^i \end{pmatrix} \sim N_{d+2}\left( \begin{pmatrix} \mathbf{X}_r^i \boldsymbol{\beta}_k \\ \mathbf{0} \end{pmatrix}, \begin{pmatrix} \mathbf{X}_r^i \boldsymbol{\Psi}_k \mathbf{X}_r^{i'} + \sigma^2 \mathbf{I}_d & \mathbf{X}_r^i \boldsymbol{\Psi}_k \\ \boldsymbol{\Psi}_k \mathbf{X}_r^{i'} & \boldsymbol{\Psi}_k \end{pmatrix} \right). \quad (7)$$

Also, from Equation (7), the posterior distribution of $\mathbf{u}_k^i$ can be written as

$$\mathbf{u}_k^i | \mathbf{y}_r^i, \boldsymbol{\beta}_k, \boldsymbol{\Psi}_k, \sigma^2 \sim N_2\left( \hat{\boldsymbol{\beta}}_k^i, \boldsymbol{\Psi}_{\hat{\boldsymbol{\beta}}_k^i} \right), \qquad (8)$$

where

$$\hat{\boldsymbol{\beta}}_k^i = (\mathbf{X}_r^{i'} \mathbf{X}_r^i + \sigma^2(\boldsymbol{\Psi}_k)^{-1})^{-1} \mathbf{X}_r^{i'}(\mathbf{y}_r^i - \mathbf{X}_r^i \boldsymbol{\beta}_k), \quad (9)$$

and

$$\boldsymbol{\Psi}_{\hat{\boldsymbol{\beta}}_k^i} = \sigma^2 \left( \mathbf{X}_r^{i'} \mathbf{X}_r^i + \sigma^2 (\boldsymbol{\Psi}_k)^{-1} \right)^{-1}. \qquad (10)$$

Figure 2(a) is a plate-like diagram that illustrates how the segment model described above generates a single waveform segment $\mathbf{y}_r^i$ when the duration $d$ of the state is given. As we enter state $s_1^i$ (that then repeats itself for $d$ time steps), the model generates the individual random effects parameter vector $\mathbf{u}_r^i$ from Equation (6), and, then, generates the observed data $\mathbf{y}_r^i = \{y_1^i, \ldots, y_d^i\}$ from Equation (5). $\mathbf{u}_k^i$ belongs to the individual waveform $i$, whereas $\boldsymbol{\Psi}$, $\boldsymbol{\beta}$, and $\sigma^2$ are global parameters.

### 3.2 Inference

To handle the random effects component in the F-B and Viterbi algorithms for segmental HMMs, we notice from Equation (7) that the marginal distribution of a segment $\mathbf{y}_r^i$ generated by state $k$ is $N_d(\mathbf{X}_r^i \boldsymbol{\beta}_k, \mathbf{X}_r^i \boldsymbol{\Psi}_k \mathbf{X}_r^{i'} + \sigma^2 \mathbf{I}_d)$, and that this corresponds to Equation (1) with the covariance matrix $\sigma^2 \mathbf{I}_d$ replaced by $(\mathbf{X}_r^i \boldsymbol{\Psi}_k \mathbf{X}_r^{i'} + \sigma^2 \mathbf{I}_d)$. Replacing the two-level segment distribution with this marginal distribution, and collapsing the hierarchy into a single level, as shown in Figure 2(b), we can use the same F-B and Viterbi algorithm as in segmental HMMs in the marginalized space over the random effects.

The F-B algorithm recursively computes the quantities in Equations (2) and (3). These are then used in the M step of the EM algorithm. The likelihood of a waveform $\mathbf{y}$, given fixed parameters $\boldsymbol{\theta} = \{\mathbf{A}, \boldsymbol{\theta}_d, \boldsymbol{\theta}_f = \{\boldsymbol{\beta}_k, \boldsymbol{\Psi}_k, (\sigma^2) | k = 1, \ldots, M\}\}$, but with states $\mathbf{s}$ and random effects $\mathbf{u}$ unknown, is evaluated as

$$\begin{aligned} p(\mathbf{y}|\boldsymbol{\theta}) &= \sum_{\mathbf{s}} \int p(\mathbf{y}, \mathbf{s}, \mathbf{u} | \boldsymbol{\theta}) \mathrm{d}\mathbf{u} \qquad (11) \\ &= \sum_{\mathbf{s}} p(\mathbf{y}, \mathbf{s} | \boldsymbol{\theta}) = \sum_k \alpha_T(k). \end{aligned}$$

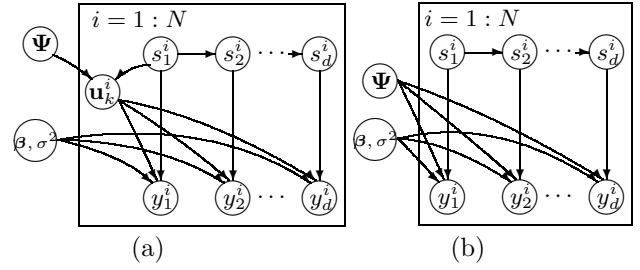

Figure 2: Plate-like diagrams for the segment distribution of random effects segmental HMMs. This shows the generative process for one segment, $y_1^i, \ldots, y_d^i$ given the duration $d$ of state $s_1^i = s_2^i = \cdots = s_d^i$. (a) shows a two-stage model with random effects parameters, and (b) the model after integrating out random effects parameters.

As in segmental HMMs, the Viterbi algorithm can be used as a method to segment a waveform by computing the most likely state sequence.

What appears to make the inference in random effects segmental HMMs computationally much more expensive than in segmental HMMs is the inversion of the $d \times d$ covariance matrix, $\mathbf{X}_r^i \boldsymbol{\Psi}_k \mathbf{X}_r^{i'} + \sigma^2 \mathbf{I}_d$, of the marginal segment distribution during the evaluation of the likelihood of a segment. For example, in the F-B algorithm, the likelihood of a segment $\mathbf{y}_r^i$ of length $d$ given state $k$, $p(\mathbf{y}_r^i | \boldsymbol{\beta}_k, \boldsymbol{\Psi}_k, \sigma^2)$, needs to be calculated for all possible durations $d$ in each of the $\alpha_t(k)$ and $\beta_t(k)$ expressions at each recursion. The naive computation of a segment likelihood using direct inversion of the $d \times d$ covariance matrix would require $O(T^3)$ computations, where $T$ is the upper bound for $d$, leading to an overall time complexity of $O(M^2 T^5)$. This can be computationally impractical when we have long waveforms with a large value of $T$, (for example, $T = 256$ for the data shown in Figure 1).

In the discussion of computational issues for random effects models, Dempster et al. (1981) suggest an expression for the likelihood that is simple to evaluate. Applying their method to the segment distribution of our model, we rewrite, using Bayes' rule, the likelihood of a segment $\mathbf{y}_r^i$ generated by state $k$ as

$$p(\mathbf{y}_r^i|\boldsymbol{\beta}_k, \boldsymbol{\Psi}_k) = \frac{p(\mathbf{y}_r^i, \mathbf{u}_k^i|\boldsymbol{\beta}_k, \boldsymbol{\Psi}_k, \sigma^2)}{p(\mathbf{u}_k^i|\mathbf{y}_r^i, \boldsymbol{\beta}_k, \boldsymbol{\Psi}_k, \sigma^2)},$$

where the numerator and the denominator of the right-hand side are given as Equations (7) and (8), respectively. The right-hand side of the above equation holds for all values of $\mathbf{u}_k^i$. By setting $\mathbf{u}_k^i$ to $\hat{\boldsymbol{\beta}}_k^i$ as given in Equation (9), we can simplify the expression for the segment likelihood to a form that involves only $O(d)$ computations for each step, where previously this involved $O(d^3)$ computations in the case of the naive approach with matrix inversions. Thus, the time complexity of the F-B and Viterbi algorithms is reduced to $O(M^2T^3)$. As shown in Mitchell et al. (1995) for segmental HMMs we can further reduce this computational complexity to $O(M^2T^2)$ by precomputing the segment likelihood and storing the values in a table—however, this precomputation is not possible with random effects models, leading to the additional factor of $T$ in the complexity term.

### 3.3 Parameter Estimation

In this section, we describe how to obtain maximum likelihood estimates of the parameters from a training set of multiple waveforms for a random effects segmental HMM using the EM algorithm. We can augment the observed waveform data with both (a) state sequences and (b) random effects parameters (both are considered to be hidden). The log likelihood of the complete data of $N$ waveforms, $D_{complete} = (\mathbf{Y}, \mathbf{S}, \mathbf{U}) = \{(\mathbf{y}^1, \mathbf{s}^1, \mathbf{u}^1), \ldots, (\mathbf{y}^N, \mathbf{s}^N, \mathbf{u}^N)\}$, where the state sequence $\mathbf{s}^i$ implies $R_{\mathbf{s}^i}$ segments in waveform $i$, is:

$$\log \ L(\boldsymbol{\theta}|D_{complete}) = \sum_{i=1}^{N} \log p(\mathbf{y}^i, \mathbf{s}^i, \mathbf{u}^i|\mathbf{A}, \boldsymbol{\theta}_d, \boldsymbol{\theta}_f)$$

$$= \sum_{i=1}^{N}\sum_{r=1}^{R_{\mathbf{s}^i}} \log P(s_r^i|s_{r-1}^i, \mathbf{A}) \tag{12}$$

$$+ \sum_{i=1}^{N}\sum_{r=1}^{R_{\mathbf{s}^i}} \log P(d_r^i|\boldsymbol{\theta}_{d_k}, k=s_r^i) \tag{13}$$

$$+ \sum_{i=1}^{N}\sum_{r=1}^{R_{\mathbf{s}^i}} \log p(\mathbf{y}_r^i|\mathbf{u}_k^i, \boldsymbol{\beta}_k, \sigma^2, k=s_r^i, d_r^i) \tag{14}$$

$$+ \sum_{i=1}^{N}\sum_{r=1}^{R_{\mathbf{s}^i}} \log p(\mathbf{u}_k^i|\boldsymbol{\Psi}_k, k=s_r). \tag{15}$$

As we can see from the above equation, given the complete data, the log likelihood decouples into four parts, where the transition matrix, the duration distribution parameters, the bottom level parameters $\boldsymbol{\beta}_k, \sigma^2$, and the top level parameters $\mathbf{u}_k^i$ of random effects models appear in each of the four terms. If we had complete data, we could optimize the four sets of parameters independently. When only parts of the data are observed, by iterating between the E step and the M step in the EM algorithm as described in the following section, we can find a solution that locally maximizes the likelihood of the observed data.

#### 3.3.1 E Step

In the E step, we find the expected log likelihood of the complete data,

$$Q(\boldsymbol{\theta}^{(t)}, \boldsymbol{\theta}) = E[\log L(\boldsymbol{\theta}|D_{complete})], \tag{16}$$

with respect to

$$p(\mathbf{S}, \mathbf{U}|\mathbf{Y}, \boldsymbol{\theta}^{(t)}) = p(\mathbf{U}|\mathbf{S}, \mathbf{Y}, \boldsymbol{\theta}^{(t)})P(\mathbf{S}|\mathbf{Y}, \boldsymbol{\theta}^{(t)})$$

$$= \prod_{i=1}^{N}\prod_{r=1}^{R_{\mathbf{s}^i}} p(\mathbf{u}_r^i|s_r^i=k, \mathbf{y}_r^i, \boldsymbol{\theta}^{(t)})P(s_r^i=k|\mathbf{y}_r^i, \boldsymbol{\theta}^{(t)}), \tag{17}$$

where $\boldsymbol{\theta}^{(t)}$ is the estimate of the parameter vector from the previous M step of the $t$th EM iteration. $P(s_r^i = k|\mathbf{y}_r^i, \boldsymbol{\theta}^{(t)})$ in Equation (17) can be obtained from the F-B algorithm. The sufficient statistics, $E\left[\mathbf{u}_k^i|s_r^i=k, \mathbf{Y}, \boldsymbol{\theta}^{(t)}\right]$ and $E\left[\mathbf{u}_k^i\mathbf{u}_k^{i'}|s_r^i=k, \mathbf{Y}, \boldsymbol{\theta}^{(t)}\right]$, for $P(\mathbf{u}_k^i|s_r^i=k, \mathbf{y}_r^i, \boldsymbol{\theta}^{(t)})$ in Equation (17) can be directly obtained from Equations (9) and (10). The computational complexity for an E step is $O(M^2T^3N)$ where $N$ is the number of waveforms.

#### 3.3.2 M Step

In the M step, we find the values of the parameters that maximize Equation (16). As we can see from Equations (12)-(15) and Equation (16), the optimization problem decouples into four parts each of which involves a distinct set of parameters (the equations are omitted here due to lack of space but are provided in Kim et al., 2004). The computational complexity for each M step is $O(MT^3N)$.

### 3.4 Model Evaluation and Score Functions

An obvious choice for evaluating a new test waveform based on a probabilistic model is to compute the likelihood of the waveform given the parameters, as shown in Equation (11). A different method that we propose in this section uses the fact that waveforms are scored based on two different aspects of how the model fits the test waveform.

Each level of the random effects model models a different source of variability. At stage two, the covariance matrix $\boldsymbol{\Psi}_k$ in Equation (6) explains the amount

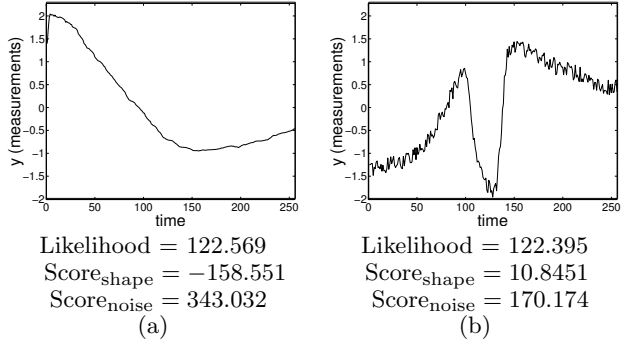

Likelihood = 122.569  
Score$_{\text{shape}}$ = −158.551  
Score$_{\text{noise}}$ = 343.032  
(a)

Likelihood = 122.395  
Score$_{\text{shape}}$ = 10.8451  
Score$_{\text{noise}}$ = 170.174  
(b)

Figure 3: Scores for test waveforms from the random effects segmental HMM trained using the data shown in Figure 1(a).

of noise in shape space. Unlike segmental HMMs, where the variance $\sigma^2$ in Equation (4) is forced to explain both shape deformations and measurement noise, random effects models allow for modelling them separately with a hierarchical structure. However, the likelihood in effect mixes both "lack of fit" terms into a single score. Consequently, smooth waveforms that are well approximated by linear segments with little measurement noise but with a considerable error in shape (as shown in Figure 3(a)) can receive the same likelihood score as waveforms with high measurement noise and little shape deformation from the mean shape (as shown in Figure 3(b)).

From the decomposition of the complete data likelihood in Equations (12)-(15), we notice that Equation (15) is a contribution from stage two of the random effects component, and that Equation (14) is a contribution from stage one. Equations (12) and (13) can be viewed as representing the shape deformation explained by the segmental HMM part of the model. The score decomposition is,

$$\text{Score}_{\text{shape}} = \text{E}[(\sum_{r=1}^{R_s} \log P(s_r|s_{r-1}, \mathbf{A}))$$
$$+(\sum_{r=1}^{R_s} \log P(d_r|\boldsymbol{\theta}_{d_k}, k = s_r))$$
$$+(\sum_{r=1}^{R_s} \log p(\mathbf{u}_k|\boldsymbol{\Psi}_k, k = s_r))|\mathbf{y}, \boldsymbol{\theta}],$$

$$\text{Score}_{\text{noise}} = \text{E}[\sum_{r=1}^{R_s} \log p(\mathbf{y}_r|\mathbf{u}_k, \boldsymbol{\beta}_k, \sigma^2, k = s_r)|\mathbf{y}, \boldsymbol{\theta}],$$

where the expectation is taken with respect to the posterior distribution of the unobserved data, $p(\mathbf{s}, \mathbf{u}|\mathbf{y}, \boldsymbol{\theta})$ (Equation (17)). Figure 3 shows examples of waveforms with these two scores. The results from our experiment in Section 4 demonstrate that using this score decomposition (i.e., using both scores as features instead of a single likelihood score) improves the recognition accuracy.

## 4 Experiments

We apply our model to two real world data sets, hot-film anemometry data in turbulent bubbly flow and ECG heartbeat data. In all of our experiments, we compare the results from our new model with those from segmental HMMs. We use several methods to evaluate the models:

**LogP Score** We compute $\log p(\mathbf{y}|\boldsymbol{\theta})$ scores (Equations (4) and (11) for each model) for test waveforms $\mathbf{y}$ to see how well the parameters $\boldsymbol{\theta}$ learned from the training data can model test waveforms.

**Segmentation Quality** To evaluate how well the model can segment test waveforms, we first obtain the segmentations of test waveforms with the Viterbi algorithm, estimate the regression coefficients $\hat{\boldsymbol{\beta}}$ of each segment, and calculate the mean squared difference between the observed data and $\mathbf{X}\hat{\boldsymbol{\beta}}$ (good segmentations produce low scores).

**One-Step-Ahead Prediction** To evaluate the predictive power of the models, we use one-step-ahead prediction on test waveforms. Given all of the observations up to time step $(t-1)$ for a test waveform $\mathbf{y}$, we compute the logP scores of the observed value at time $t$ and the mean squared error of the predicted values, for the next time step.

In all of these experiments, we use five-fold cross validation. To evaluate the performance of our model for classification, we include in the test set negative examples of the shape that we are modeling, and build a $k$-nearest neighbor classifier with varying values for $k$ using the scores from the model as a feature vector for each waveform. For the model from each of the five-fold cross validation runs, using the positive examples in the test set for that model and the negative examples, we use three-fold cross validation to obtain the classification accuracy of the classifier.

### 4.1 Hot-film Anemometry in Turbulent Bubbly Flow

Hot-film anemometry is a technique commonly used in turbulent bubbly flow measurements in fluid physics. Interactions between the bubbles and the probe in turbulent gas flow, such as splitting, bouncing, and penetration, lead to characteristic interaction waveform shapes. Physicists are interested in detecting the occurrence and type of interactions automatically from such waveforms (Bruun, 1995). There can be large variability in the shape of the waveforms caused by various factors such as velocity fluctuations and different gas fractions during measurement. Labels are available for the type or class of each interaction based on high-speed image recordings of the event obtained simultaneously with the interaction signal. In the results of this paper, we model waveforms for one specific type of interaction where the probe splits the bubble. Our data consist of 50 waveforms such as those shown in Figures 1(a) and (b). We randomly sampled 20 waveforms from this data set to form a training set for each of five-fold cross validation runs. Given that Figure 1(c) is a reasonable piecewise linear approximation of the general shape, we subjectively chose $M = 6$ as the number of states for both segmental HMMs and random effects segmental HMMs.

Table 1: Performance on Bubble-probe Interaction Data

|  | LogP Scores | One Step Ahead Prediction | | Segmentation Error |
|---|---|---|---|---|
|  |  | LogP | Mean Squared Error |  |
| Segmental HMMs | -75.92 | -0.2824 | 0.1035 | 0.0231 |
| Random Effects Segmental HMMs | 248.68 | 0.9863 | 0.0247 | 0.0050 |

Table 2: Performance on ECG Data for Normal Heartbeats

|  | LogP Scores | One Step Ahead Prediction | | Segmentation Error |
|---|---|---|---|---|
|  |  | LogP | Mean Squared Error |  |
| Segmental HMMs | 64.59 | 0.2073 | 0.0630 | 0.00620 |
| Random Effects Segmental HMMs | 394.71 | 1.9393 | 0.0068 | 0.00052 |

Figure 4(a) illustrates visually that the quality of the segmentations of the waveforms using the Viterbi algorithm is much better with random effects than without. Table 1 shows a reduction of approximately 80% in squared error from using random effects for these segmentations. Table 1 also shows a significant increase in logP scores for the test waveforms in the models with random effects parameters as well as significantly better one-step-ahead predictions. To evaluate the performance of the models for classification, we used 72 additional waveforms of negative examples labeled as bouncing, penetrating, and glancing interaction types, and plot the classification accuracy in Figure 5. In addition to the two probabilistic models, we include the results of using the direct mean squared distance between two waveforms as a distance measure in $k$-nearest neighbor algorithms (as a baseline method). Using the two decomposed scores improves the accuracy of $k$-nearest neighbor classifiers significantly over just using the likelihood.

### 4.2 ECG Data

The shape of heartbeat cycles in ECG data can be used to diagnose the heart condition of a patient (Koski, 1996; Hughes et al. 2003). For example, Figure 4(b) shows the typical shape of normal heartbeats, whereas Figure 4(c) is taken from a heart experiencing a premature ventricular contraction. However, even among heartbeat recordings for the same heart condition from the same individual, there is a significant variability in terms of shape and length. We chose an ECG recording from the MIT-BIH Arrhythmia database, and manually divided it into individual waveforms to obtain 28 normal heartbeats and 28 abnormal heartbeats of a premature ventricular contraction. 10 waveforms from each of the resulting data sets were used to train the models with the number of states $M = 9$ in normal cases and $M = 6$ in abnormal cases. The results from 5-fold cross validation are shown in Table 2 and 3. Again we see a significant improvements for the random effects model. In terms of classification (details not shown) our new models were 100% accurate in all experiments versus an average accuracy of 98% for segmental HMMs.

## 5 Conclusions

In this paper, we proposed a probabilistic model that extends segmental HMMs to include random effects. This model allows an individual waveform to vary its shape in a constrained manner via a prior distribution over individual waveform parameters. We demonstrated that random effects segmental HMMs can achieve a significant improvement in modeling, segmentation, and classification of waveforms.

### Acknowledgements

This material is based upon work supported by the National Science Foundation under Grant No. SCI-0225642 for the OptIPuter project. We would like to

Table 3: Performance on ECG Data for Premature Ventricular Contractions

| | LogP Scores | One-Step-Ahead Prediction | | Segmentation Error |
|---|---|---|---|---|
| | | LogP | Mean Squared Error | |
| Segmental HMMs | 28.82 | -0.1738 | 0.0550 | 0.0075 |
| Random Effects Segmental HMMs | 323.79 | 1.5179 | 0.0141 | 0.0014 |

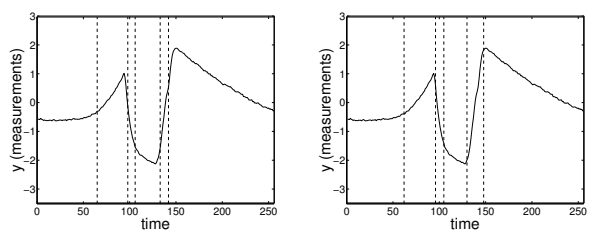

(a) Bubble-probe interaction data

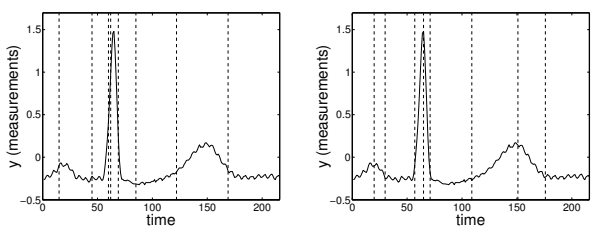

(b) ECG data - normal heartbeats

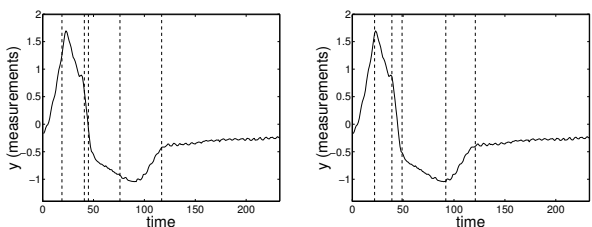

(c) ECG data - premature ventricular contractions

Figure 4: Examples for the segmentation of waveforms by the Viterbi algorithm for segmental HMMs (left) and for random effects segmental HMMs (right)

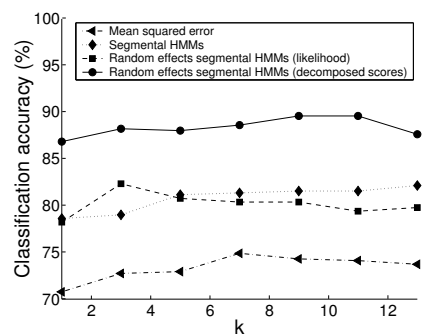

Figure 5: $k$-nearest neighbor classification accuracy

thank David Van Dyk for discussions relating to random effects models and EM.